\documentclass[12pt]{article}  
\usepackage{amsfonts,xypic}   
\usepackage{mathptmx} 

\usepackage[english]{babel}

\begin{document}

\title{Cosmological Constant $\Lambda$ \emph{vs.} Massive Gravitons: \\ A Case Study in 
 General Relativity Exceptionalism \emph{vs.} Particle Physics Egalitarianism} 
 
\author{J. Brian Pitts \\ Faculty of Philosophy  \\ University of Cambridge }

\date{23 July 2019} 

\maketitle

\abstract{The renaissance of General Relativity witnessed considerable progress regarding both understanding and justifying Einstein's equations. Both general relativists and historians of the subject tend to share a view, General Relativity exceptionalism, which emphasizes how General Relativity is novel and unlike the other fundamental interactions. But does some of the renaissance progress in understanding and justifying Einstein's equations owe something to an alien source, namely particle physics egalitarianism, which emphasizes how similar General Relativity is to the other fundamental interactions? If so, how should the historiography of gravitation and Einstein's equations reflect that fact? Two areas of renaissance progress can be considered very briefly (gravitational waves and the question of energy, and the justification of Einstein's field equations in terms of particle physics) and one in detail:  the longstanding confusion over the relation between Einstein's cosmological constant $\Lambda$ and a graviton mass, confusion  introduced in 1917 with $\Lambda.$ 

	Regarding gravitational waves, General Relativity exceptionalism discouraged trusting the linear approximation, in which the existence of gravitational radiation is evident, whereas particle physics egalitarianism encouraged such trust. While general relativists differed on this issue in the  1950s, the rare particle physicist seriously involved, Feynman, was among those who invented the sticky bead argument that largely resolved the controversy. 
Regarding the justification of Einstein's equations in terms of particle physics, the spin $2$ derivations of Einstein's equations provided considerably support of the view that Einstein's equations correctly describe gravitation.

	The idea of a graviton mass has a 19th century Newtonian pre-history in Neumann's and Seeliger's long-distance modification of gravity, which (especially for Neumann) altered Poisson's equation to give  a potential $e^{-mr}/r$ for a point mass, improving convergence for homogeneous matter. Einstein reinvented the idea before introducing his faulty  analogy with $\Lambda$. This confusion was first critiqued by Heckmann in the 1940s (without effect) and by Trautman,  DeWitt, Treder, Rindler, and Freund \emph{et al.} in the 1960s, and especially more recently by Sch\"{u}cking, but it has misled North, Jammer, Pais, Kerszberg, the Einstein Papers, and Kragh. The error is difficult to catch if one has an aversion to perturbative thinking, but difficult to make if one thinks along the lines of particle physics. $\Lambda$ contributes predominantly a zeroth order term to the field equations (a constant source), whereas a graviton mass contributes a linear term. Nonperturbatively, massive spin 2 gravity is bimetric. The $\Lambda$-graviton mass confusion not only distorted the interpretation of Einstein's theory, but also obscured a potentially serious particle physics-motivated rivalry (massless \emph{vs.} massive spin 2). How could one entertain massive spin 2 gravity if $\Lambda$ is thought already analogous to the Neumann-Seeliger scalar theory?  Massive spin 2 gravity would encounter problems within particle physics in the early 1970s, which have been substantially resolved in recent years (though new problems have arisen).  Other authors manage to avoid confusing $\Lambda$ with the Neumann-Seeliger long-range decay but fail to recognize that the latter, with the physical meaning of a graviton mass, provides an interesting case of unconceived alternatives.  

	In sum, both the interpretation and the justification of Einstein's equations owed some of their renaissance progress to particle physics egalitarianism. Historiography, like physics, is best served by overcoming the divide between the two views of gravitation.}



\section{Introduction}                      


It is often noticed  that there is a great gulf  between the views of gravitation held by general relativists and those held by particle physicists \cite{KaiserGRParticle,Rovelli,BrinkDeserSupergravity,BlumLalliRennIsis}.   These  communities   have different expectations for how gravity, especially Einstein's General Relativity, should fit into physics:  general relativists tend to incline toward an exceptionalist position that expects General Relativity to be novel, dramatic, and non-perturbative, whereas particle physicists incline toward an egalitarian view that expects General Relativity to be familiar and perturbative, as just another field (in this case mass $0$ and spin $2$) to be quantized.  As Feynman describes his approach to gravitation in his lectures on gravitation, 
\begin{quote}
Our pedagogical approach  is more suited to meson theorists who
have gotten used to the idea of fields, so that it is not hard for them to conceive that
the universe is made up of twenty-nine or thirty-one other fields all in one grand
equation; the phenomena of gravitation add another such field to the pot, it is a
new field which was left out of previous considerations, and it is only one of the
thirty or so; explaining gravitation therefore amounts to explaining three percent
of the total number of known fields.
\cite[p. 2]{Feynman} \cite{KaiserGRParticle} \end{quote}
 For present purposes, another relevant feature of the mental training of Feynman's theorist of mesons and nucleons is a habit of contemplating mass terms:  the default relativistic wave equation for the particle physicist is the Klein-Gordon equation, not the relativistic wave equation, which is merely the special case when the field/particle in question is massless.  It has been known since the 1930s that massless fields/particles are naturally  associated with gauge freedom for spins greater than or equal to $1$ (vectors, vector-spinors, symmetric tensors, \emph{etc.}) \cite{Fierz}.  There is also a presumption of a smooth massless limit \cite{UnderdeterminationPhoton}.  This presumption, however, can be defeated \cite{Zakharov,vDVmass1,vDVmass2}, and this defeat apparently can be overcome non-perturbatively \cite{Vainshtein,Vainshtein2}, a topic of much literature in the 2000s. Mass terms also make a significant  difference conceptually  even for the scalar case where only a smaller symmetry group, not gauge freedom, is involved \cite{ScalarGravityPhil}.
These differing  attitudes toward gravity are reflected in the quantization programs pursued in both camps.  

 General relativists are also far more inclined toward philosophical and historical considerations than are particle physicists.  It is therefore not very surprising that historical and philosophical work on 20th century gravitation theory would usually  favor the viewpoint of general relativists, consciously or otherwise.    If gravitational physics is best pursued by ``amphibious'' physicists \cite{TenerifeProgressGravity} (of whom Stanley Deser and the late Bryce DeWitt  would be prominent examples), what would the history and philosophy of gravitational physics from an amphibious perspective look like?  

 There are examples where a particle physics-informed history and philosophy of space-time and  gravitation theory would have new insights, diagnosing missed opportunities and poor arguments, offering new alternatives and arguments, and  suggesting a different heuristic based on alternate expectations about how General Relativity fits into physics.  Ideally the physical insight of both physics communities would be systematically integrated.  One would also hope that scientific progress will gradually test each physics community's expectations in some definitive fashion, perhaps with mathematical calculations and/or computer simulations.

The 1950s-70s witnessed a renaissance of General Relativity \cite{BlumLalliRennIsis,BlumLalliRennGRRenaissance,BlumGiuliniLalliRennEPJHintro,BlumLalliRennWavesLongRevolution}.  Prior to that time, the clearly dominant  party  in theoretical physics was particle physics. General Relativity, when studied at all, tended to appear in departments of mathematics, and Einstein's 1915-16 equations were not necessarily viewed as the final destination.  Among particle physicists, efforts were made to quantize Einstein's equations  on  the model of other field theories.  But little emphasis on  novel conceptual lessons about space-time could arise along such lines.  One might be tempted to think that such particle physics egalitarianism was an aspect of the decades of stagnation for General Relativity, and that the renaissance of General Relativity partly consisted in overcoming such particle physics attitudes.  There is, of course, some degree of normative coloration here, one way or the other:  while a ``renaissance'' is supposed to be good, the goodness of an historical tendency to emphasize the distinctiveness of General Relativity depends in part on the normative question of how distinctive General Relativity in fact is, a topic that remains controversial in physics.  If one were to compare the largely sound treatment of $4$-dimensional symmetry in the Hamiltonian formulation of General Relativity by Rosenfeld and by Anderson and Bergmann \cite{RosenfeldQG,SalisburySundermeyerRosenfeldQG,AndersonBergmann} with the problem of missing change (observables as constants of the motion) in the mid-1950s \cite{BergmannGoldberg} and the  disappearance of $4$-dimensional symmetry by the late 1950s \cite{DiracHamGR}, one would quickly encounter topics of current research interest in physics \cite{ObservablesEquivalentCQG} and might  less confidently expect  revolutionary  novelty to be helpful.  %

This paper  briefly notes two  areas where the renaissance of General Relativity saw progress that was \emph{positively} related to the particle physicists' egalitarian attitude  and then mostly attends to one more.  Of the two areas merely noted, one is gravitational waves:  whether they exist and whether they carry energy.  This topic has been well treated  \cite{KennefickWaves,Cattani}, albeit without reference to particle physics.  It might be useful, with a view to Feynman's contribution regarding the sticky bead argument, to consider how a particle physics egalitarian perspective  was helpful in this context.   Feynman  identified ``a perennial prejudice that gravitation is somehow mysterious and different'' as why General Relativists ``feel that it might be that gravity waves carry no energy at all.'' \cite[p. 219]{Feynman} 

A second area of renaissance progress for General Relativity that could be discussed involves particle physicists' ``spin $2$'' \emph{derivation} of Einstein's equations \cite{Gupta,Kraichnan,Deser,SliBimGRG,ConverseHilbertian}. There is a large difference between a pedestrian re-interpretation of General Relativity in terms of flat space-time and universal distortion forces (which might be viewed as an effort to miss the point of the theory by forcing it into a more familiar mold) and a \emph{derivation} of Einstein's equations assuming flat space-time premises; the difference is especially strong regarding what justifies attending to Einstein's field equations rather than some other field equations.   While these derivations admittedly sometimes have been motivated by or invoked in support of egalitarian or conservative/anti-revolutionary sentiments, the fact remains that a demonstration that one can hardly avoid Einstein's equations, given particle physicists' own criteria such as avoiding explosive instabilities \cite{VanN} (criteria far less negotiable than those of general relativists such as Einstein's  involving rotation \cite{JanssenNemesis,EinsteinEnergyStability}), added considerably to the justification of Einstein's equations.  What would have happened if particle physics had been seen as motivating rivals to Einstein's equations?   Due to limitations of space, this topic will not be discussed further here.  
 
The last area of renaissance progress, to which this chapter mostly attends,  involves substantial clarification of Einstein's 1917  confusion  between a graviton mass (or the ancestor thereof) and Einstein's cosmological constant $\Lambda$.  Removing this confusion both makes  conceptual space for massive spin $2$ gravitational theories (which encountered new problems in the early 1970s \cite{vDVmass1,vDVmass2,DeserMass} before the 2000s-10s revival) and clarifies the meaning of $\Lambda$. 
Hence the meaning of Einstein's equations becomes clearer (even if the justification of his equations potentially becomes weaker due to the conceptualization of a serious rival in the form of massive spin $2$ gravity  \cite{UnderdeterminationPhoton}).   Thus, perhaps ironically given the real or supposed opposition between general relativist and particle physicist viewpoints, particle physics contributed to the renaissance of General Relativity in several ways.  The goal of an amphibious approach, of course, is not to replace one kind of partisanship with another kind, nor even to achieve a balance of partisanships, but to overcome the physical divide and  do the history and philosophy of gravitation and space-time theory  using the whole body of physical knowledge.

Unfortunately this amphibious enterprise  has had so little presence historically that a polemical attitude will be inevitable toward some traditional work.  While many of the authors singled out for getting the issue right are physicists, initially they are writing to correct the errors of other physicists, whether past (Einstein in the 1910s) or more current (in the 1960s).  Some of the authors criticized on the historical side, \emph{viz.} Pais and Jammer, were also physicists.  They get classed in history not due to any  limitations in physics participation, but due to their extensive work and achievements in the history of science and for the purposes of their works considered here.  It is inevitable that research physics literature will be somewhat `ahead' of historical literature.  However, the generally commendable historical emphasis on original sources, context, \emph{etc.}, in this case has the effect of perpetuating Einstein's mistake.  In this regard, the Einstein Papers (volume 6)  \cite[p. 552]{EinsteinCosmologicalGerman} gave Einstein's faulty analogy  renewed vigor by providing commentary that did not notice the corrections that had been made over the years and by citing North's flawed treatment \cite{North} (on which more below) for further discussion.



\section{Cosmological Constant \emph{vs.} Graviton Mass:  A Recurring Confusion} 

\subsection{Historical Background} 

In introducing his cosmological constant $\Lambda$ to the world \cite{EinsteinCosmologicalGerman}, Einstein claimed that $\Lambda$  was analogous to a long-range modification of the Poisson equation that, as a matter of fact, produces a faster (exponential) decay:
\begin{quote} 
\ldots  the system of equations [that are his field equations]  allows a readily suggested extension which is compatible with the relativity postulate, and is perfectly analogous to the extension of Poisson's equation given by equation (2).
\cite{EinsteinCosmologicalGerman} \end{quote} 
This analogy, unfortunately, is incorrect.  Below it will appear that, after criticism in the 1940s that had negligible effect, substantial criticism from a number of noteworthy physicists  appeared in the 1960s.  It will also appear that historians and philosophers of physics --- there seems to be little useful distinction between historians and philosophers in this context --- continued to accept Einstein's faulty analogy for decades thereafter, though in the last two decades such confusion has become considerably rarer.

  The idea of a graviton mass is due to 1920s work on relativistic wave equations, especially the Klein-Gordon equation, 1930s work on the Yukawa potential, and the 1930s recognition that gravity, even according to Einstein's equations, could be construed as occupying a well-defined place in the taxonomy of relativistic (at least Poincar\'{e}-invariant) wave equations \cite{FierzPauli,Fierz2}. If one expects every field theory to be quantized and to yield ``particles'' much as quantizing Maxwell's electromagnetism yields massless photons and quantizing de Broglie-Proca massive electromagnetism (with an additional term $ -\frac{1}{2} m^2 A^{\mu}  A_{\mu}$ in the Lagrangian density) yields massive photons, then quantizing Einstein's theory ought to yield massless ``gravitons'' and quantizing a related theory with a suitable term quadratic in the gravitational potentials ought to yield massive gravitons.  
Notwithstanding the quantum words and promissory notes about particles, the basic idea is just classical field theory and partial differential equations. A ``mass'' in effect is an inverse length scale (the conversion being effected using $c$ and $\hbar$). 

 There  is thus a significant pre-history of massive gravitons from the late 19th century.  
 That is due especially to Neumann's and Seeliger's modification of Newtonian gravity in the 1890s with an exponentially decaying potential \cite{PockelsHelmholtzEquation,Neumann,Neumann1886,Seeliger1896,Neumann1902Finite,Pauli,North,NortonWoes}. Neumann paid considerable attention to this differential equation, whereas Seeliger tended to modify the force rather than the potential, at times using a point mass force law of $e^{-\lambda r}/{r^2}$  \cite{Seeliger1895a}.  
Seeliger provided, if not a physical meaning (as the graviton mass later would), at least a physical motivation, namely, rendering convergent various integrals that misbehaved for Newtonian gravity with an infinite homogeneous matter distribution.  
Later  Einstein in his popular treatment  \cite[pp. 362, 363 of the translation]{EinsteinSpecial}  
offered some criticisms of  Seeliger's modification of the Newtonian force law:
\begin{quote} 
Of course we purchase our emancipation from the fundamental
difficulties mentioned, at the cost of a modification and complication
of Newton's law which has neither empirical nor 
theoretical foundation. We can imagine innumerable laws
which would serve the same purpose, without our being able
to state a reason why one of them is to be preferred to the
others; for any one of these laws would be founded just as
little on more general theoretical principles as is the law of
Newton.
\end{quote} 
 But the uniqueness problem does not hold in light of Neumann's mathematics (which Einstein reinvented), while the Klein-Gordon equation and Yukawa potential would later give a physical  meaning to Neumann's mathematics.  The empirical basis, while not empty, is merely analogous:  many fields are massive, so why not the graviton field also?      
In the actual contingent history, Einstein was unaware of Seeliger's work until  after the `final' 1915-16 field equations were known  \cite[p. 420]{EinsteinSpecial}  \cite[p. 557]{EinsteinForster} \cite{EinsteinLunarGerman} \cite[p. 189]{EinsteinNotes} (\emph{c.f.} \cite{EarmanLambda}).

%

%

As early as 1913 Einstein enunciated a principle to the effect that the field equations for gravity should not depend on the absolute value of the gravitational potential(s) 
 \cite[p. 465]{NortonNordstromBook} \cite[pp. 544, 545]{EinsteinPresent}.  It follows, given the concept of a mass term, that a mass term is not permitted, because mass terms make the field equations contain the potential(s) algebraically. (The cosmological constant provides a subtle way of maintaining gauge freedom with an algebraic term, but does not permit a graviton mass.)   It is evident that historians of General Relativity with an eye for particle physics would be more apt to recognize how Einstein's principle generates a problem of unconceived alternatives.


\subsection{Physical Background}

It isn't initially obvious what connection there might be, if any, between a graviton mass and the cosmological constant $\Lambda.$  A graviton mass involves a quadratic term in the Lagrangian like $ -\frac{1}{2}  m^2 A_{\mu} A^{\mu}$ in the electromagnetic case, which by differentiation implies a linear term in $A_{\mu}$ in the Euler-Lagrange equations.  One might thus envisage something like  $ -\frac{1}{2}  m^2 \gamma_{\mu\nu} \gamma^{\mu\nu}$ for gravity. Whether or not one actually quantizes gravity, one can follow particle physics custom and refer to such a term as a graviton mass term.  (No other terminology is available.  Indices are assumed to be raised or lowered with a background metric $\eta_{\mu\nu}$. The simplest choice is a flat metric.)  Massive electromagnetism was first entertained by de Broglie \cite{deBroglieBlack,deBrogliePhilMag} and developed as a field theory by Proca \cite{Proca}.  Massive gravity was also encouraged by de Broglie and was pursued in France during the 1940s onward  \cite{TonnelatWaves,Tonnelat20,TonnelatGravitation,PetiauCR44a,PetiauRadium45,DrozVincentMass59}, and later also in Sweden  \cite{BrulinHjalmarsSpin2Graviton};   in both countries spinor rather than tensor notation was often used (except by Droz-Vincent).  
 Recognizably modern work appeared in the mid-1960s \cite{OP,FMS}; the former paper is still valuable for its radical conceptual innovations (such as inventing nonlinear group realizations) achieved with  binomial series expansions to take arbitrary powers of the metric tensor (expanded about the identity matrix using $x^4 = ict$!).  Thus they  invented many theories of gravity (including some recently reinvented) and showed how to Ockhamize the coupling of gravity to spinors as well \cite{OPspinor,PittsSpinor} (previously partly anticipated by Bryce DeWitt's series expansions \cite{DeWittSpinor,DeWittDissertation} and  still not widely known outside the supergravity community).  Some of these results on the symmetric square root of the metric and massive gravity theories using it were reinvented in the 2010s.

Complications arise, however, with a symmetric tensor potential.  A first complication is that with two indices, $\gamma_{\mu\nu}$ admits a trace $\gamma =_{def} \gamma_{\mu}^{\mu},$ which cannot justly be ignored as a participant in the mass term, so there will be a new coefficient for the new scalar term involving $m^2 \gamma^2.$  A second complication (partly following from the first) is that, unlike electromagnetism, gravity admits many (indeed infinitely many) relevant distinct but comparably plausible definitions of the gravitational potential.  Comparing two such definitions, such as $g_{\mu\nu} -\eta_{\mu\nu}$ and $-\mathfrak{g}^{\mu\nu} + \sqrt{-\eta}\eta^{\mu\nu},$ one differs from another by how much of the trace term is mixed in and what nonlinearities are included.  There is no `correct' answer, although there are few incorrect ones, for which the relation to the others cannot be inverted.  Physical meaning of an expression such as $\gamma_{\mu\nu}$ is achieved by relating $\gamma_{\mu\nu}$  to the effective curved metric $g_{\mu\nu}$ and  $\eta_{\mu\nu}$.  A third complication  is that the trace $\gamma$ suggests a negative-energy ``ghost'' degree of freedom, which is likely to lead to explosive instability in quantum field theory.  One can tune away this ghost at linear order \cite{FierzPauli}.  But such a theory seems not to have the expected massless limit of General Relativity \cite{vDVmass1,Zakharov,Iwasaki,vDVmass2},  the van Dam-Veltman-Zakharov discontinuity, making pure spin $2$ massive gravity apparently refuted by experimental data.  
  A fourth complication is that even if one tunes away the ghost at lowest order, it reappears nonlinearly \cite{TyutinMass,DeserMass}, the Boulware-Deser ghost.  The apparent dilemma of empirical falsification or explosive instability largely stopped research on massive gravity from \emph{c.} 1972-1999.\footnote{The obvious exception was the Russian school of A. A. Logunov and collaborators starting in the late 1970s, which was largely ignored by others or occasionally subject to polemics \cite{Zel2}, not without some justification. Logunov being the editor of the Russian original of \emph{Theoretical and Mathematical Physics} and a Soviet and Russian Academician, he was able to maintain a noticeable research group with many publications.  }   The apparent empirical need for a cosmological constant  \cite{Riess,Perlmutter}, however, undermined confidence in General Relativity especially on long distance scales, thus cracking open the door for renewed consideration of massive gravity.  The van Dam-Veltman-Zakharov discontinuity was plausibly resolved during the 2000s by a non-perturbative treatment called the Vainshtein mechanism, which built on an early suggestion by Arkady Vainshtein   \cite{Vainshtein,Vainshtein2}.  The Boulware-Deser ghost was  resolved in the 2010s \cite{deRhamGabadadze,HassanRosen}  (but see the neglected early work of Maheshwari \cite[Appendix]{MaheshwariIdentity,MassiveGravity3,FMS}). 
%
Thus massive gravity, not necessarily in this simple form, is now a lively field of research overcoming the GR-particle physics split, spawning review articles in prestigious places \cite{deRhamLRR,HinterbichlerRMP} and getting some of its (re-)developers good physics jobs.   After decades of darkness, massive gravity is very much a part of the current physics scene.  These decades of darkness, however, have left their mark on the historical and philosophical work on gravity, in that until recently one would have had to look in unusual places to acquire any knowledge of such matters.

How one can compare an expression like $\Lambda \sqrt{-g}$ to a mass term at all?  $\sqrt{-g}$ seems not to be any kind of series and is not expressed in terms of a graviton potential $\gamma_{\mu\nu}$ that vanishes when `nothing is happening.' Rather, the trivial value for $\sqrt{-g}$ is $1,$ not $0$.  (This over-simple claim is coordinate-dependent  but heuristically useful.)  It also isn't clear how differentiating $\Lambda \sqrt{-g}$ to find the Euler-Lagrange equations leads to something one order lower in the potential, as one might have expected; indeed the expression $\frac{\partial \sqrt{-g} }{\partial g_{\mu\nu}}= \frac{1}{2} \sqrt{-g} g^{\mu\nu}$ is of order $1$ (dominated by a zeroth order term in the potential, so to speak), just as $\sqrt{-g}$ is.  One can render $\sqrt{-g}$ comparable to a graviton mass expression (quadratic in $\gamma_{\mu\nu}$) using a perturbative expansion of  $g_{\mu\nu}$ about  $\eta_{\mu\nu},$  defining  $\gamma_{\mu\nu}$ by $g_{\mu\nu} = \eta_{\mu\nu} + \sqrt{32 \pi G} \gamma_{\mu\nu},$ though suppressing the normalizing $\sqrt{32\pi G}$ is sometimes clarifying and is employed here.  (Infinitely many other choices of field variables are possible \cite{OP,MassiveGravity2,MassiveGravity3}, leading to differences of detail that can be important in some contexts.)  The determinant $g$ becomes a quartic polynomial in $\gamma_{\mu\nu}$ (though quadratic terms suffice for present purposes):
$$ g = \eta (1 + \gamma  - \frac{1}{2} \gamma_{\mu\nu}\gamma^{\mu\nu} + \frac{1}{2} \gamma^2 + O(\gamma^3) + O(\gamma^4)).$$ 
 One can find $\sqrt{-g}$ using the binomial series expansion (here with $n = \frac{1}{2}$ and using  an obvious formal extension of the factorial notation for convenience)
$$ (1+ x)^n = \sum_{i=0}^{\infty}   \frac{n!}{(n-i)! i!}  x^i  =  \sum_{i=0}^{\infty}   \frac{ n \cdot (n-1)  \ldots  (n-i+1)}{i!} x^i = 1 + nx + n(n-1)x^2/2\ldots.$$  
The coefficient of $\gamma^2$ gets contributions from two different terms. 
Thus  $$\sqrt{-g} = \sqrt{-\eta}(1 + \frac{1}{2} \gamma  - \frac{1}{4} \gamma_{\mu\nu} \gamma^{\mu\nu} + \frac{1}{8} \gamma^2 + \ldots):$$  the cosmological constant term in the Lagrangian density is an infinite series of powers of the graviton potential, starting with zeroth order.  Thus its contribution to the field equations is also such an infinite series.  While the zeroth order term in the Lagrangian density does not influence the field equations, the first-order term (perforce a constant times $\gamma$) is all-important in giving the characteristic $\Lambda$ phenomenology arising from a constant term in the field equations.   
There are, of course, other ways of having a linear term in the Lagrangian and hence a constant in the field equations:  one could simply install a term of the form $\sqrt{-\eta} (g_{\mu\nu} \eta^{\mu\nu} -4)$, but there is little motivation for such a term in isolation.    The cosmological constant, by contrast, provides a motivation for such a term.  The  precise tuning of linear, quadratic and higher terms in the cosmological constant term preserves general covariance (in the sense of admitting arbitrary coordinates with only fields varied in the action and no gauge compensation fields \cite{PittsArtificial,FriedmanJones}).  To have instead a graviton mass term rather than a cosmological constant, one needs to remove the linear term from the Lagrangian and hence the zeroth order term from the field equations, leaving a quadratic term in the Lagrangian and hence a linear term in the field equations.  Some old  \cite{OP,FMS,MaheshwariIdentity} and recent  \cite{HassanRosen} works on massive gravity  therefore use expressions along the lines of $\sim \sqrt{-g} + \sim g_{\mu\nu} \eta^{\mu\nu} \sqrt{-\eta} + \sim \sqrt{-\eta},$ where the coefficients are chosen to cancel the linear term in the Lagrangian (to avoid $\Lambda$ phenomenology) and the zeroth order term (for the tidiness of having the Lagrangian density vanish when the graviton potential does). The details of the middle term admit considerable variety such as some constant times  $g_{\mu\nu} \eta^{\mu\nu} \sqrt{-\eta}$, or $\sqrt{-g}^{84.6} g^{\mu\nu} \eta_{\mu\nu} \sqrt{-\eta}^{-83.6}$, or non-rational density weights or even non-rational powers of the metrics \cite{OP}.  Such generality was  reinvented recently \cite{FMS,MaheshwariIdentity,deRhamGabadadze,HassanRosen,MassiveGravity3}  in order to (re)discover nonlinearly ghost-free massive gravities (that is, theories such that there are no negative-energy degrees of freedom even when nonlinear terms are considered).

 A key point is that having a mass term requires two metrics, and consequently the gauge freedom (substantive general covariance) of General Relativity is removed, leaving more degrees of freedom.  
It is, in fact, possible to have both a cosmological constant and a graviton mass if there are linear and quadratic terms in the Lagrangian density, but their coefficients are not related as in $\sqrt{-g}.$  Cubic and higher terms can be construed as interactions and are not very important empirically in comparison to the linear and quadratic terms.

How does the cosmological constant $\Lambda$ differ from a graviton mass term in its effects on the equations of motion of test particles? 
The geodesic equation of motion for a point test particle, assuming slow motions, weak fields, nearly Cartesian coordinates,  and $-+++$ signature, has 
spatial components $\frac{d^2 x^i}{dt^2} - \frac{1}{2} \partial_i g_{00} =0.$ Identifying this approximate result with the Newtonian result $\frac{d^2 x^i}{dt^2} = - \partial_i \phi$ for the Newtonian potential $\phi,$ one  obtains $g_{00} \approx -c^2 -2\phi.$ 
%
%
Because of the constant term $-c^2,$  the cosmological constant leads at lowest order to a  \emph{zeroth} order term in the field equations, not to the antecedently more physically plausible 19th century modification with a linear algebraic term
$$\nabla^2 \phi - \lambda \phi = 4 \pi G \rho.$$  
One can assess the relative sizes of the terms such as  $-c^2$ and $-2  \phi $ using Newton's constant $G= 6.674\cdot 10^{-11} N\cdot m^2 kg^{-2},$ the masses of the Sun and the Earth, the radius of the Earth's orbit, the radius of the Earth, and the speed of light.  Taking $\phi \approx - \frac{GM}{r}$ (which will hold approximately for equations approximating the Poisson equation for Newtonian gravity), one has for the potential from the Sun at the Earth $$\phi \approx - (1.99 \cdot 10^{30}kg) \cdot (6.674\cdot 10^{-11} N\cdot m^2 kg^{-2}) / (1.50 \cdot 10^6 km) \approx - 8.85 \cdot 10^9 \frac{m^2}{s^2}.$$  The potential from the Earth at its surface is analogously $$\phi \approx - (5.97 \cdot 10^{24} kg) \cdot  (6.674\cdot 10^{-11} N\cdot m^2 kg^{-2}) / 6371km \approx -6.25 \cdot 10^{7} \frac{m^2}{s^2}.$$  By contrast $c^2 \approx 9\cdot 10^{16} \frac{m^2}{s^2}.$  Thus typical values of the potential for both terrestrial and solar system effects are vastly smaller (in absolute value) than $c^2$. 
The strange \cite{FMS,Schucking}  zeroth order term  will tend to dominate over the intended  linear  term in $\phi$.  The potential grows quadratically and, if $\Lambda >0,$ is repulsive in Einstein's theory, giving an anti-oscillator force (proportional to distance like a spring, but with the `wrong' sign).  By contrast a graviton mass term leaves the gravitational force attractive, but merely makes it decay faster than $\frac{1}{r^2}$ at long distances due to exponential decay.\footnote{Mathematically with the graviton mass term one has a superposition of exponentially decaying and exponentially growing factors times $\frac{1}{r},$ but one routinely discards the growing solution on grounds of physical reasonableness.  With the zeroth order term in the field equations, by contrast, there are no solutions to spare and a (quadratically) growing solution cannot be discarded.}   
Eddington, without comparing $\Lambda$ to a graviton mass (or Neumann's antecedent) or criticizing $\Lambda$ as bizarre,  described the phenomena fairly adequately in 1923 \cite{Schucking}:
\begin{quote}
Hence \hspace{1in}  $\frac{d^2r}{ds^2} = \frac{1}{3} \lambda r$...................(70.22).

Thus a particle at rest will not remain at rest unless it is at the origin; but will be repelled from the origin with an acceleration increasing with distance.
\cite[p. 161]{Eddington}  \end{quote}
It is easy to imagine that such a description would help to draw attention to the difference between $\Lambda$ and the idea of a graviton mass, which would become easier and easier to conceive during the 1920s and '30s.  But  such a result seems not to have occurred.  
A subtler mistake that arose during  the 1960s held  that the cosmological constant was exactly analogous to a Neumann and Seeliger-type long-modification of gravity, if  not for \emph{static} fields, at least  for  gravitational wave propagation. %



Since the 1910s an exact solution for Einstein's equations with spherical symmetry and a cosmological constant $\Lambda$ has been known due to Kottler and  Weyl \cite{KottlerSchwarzschilddeSitter,WeylSchwarzschilddeSitter,Perlick,Ellis2Mass}  \cite[pp. 177, 397]{Ohanian}.  It is now known among physicists (with naming conventions more suited to economy by taking equivalence classes under coordinate transformations than to historical accuracy) as the Schwarzschild-de Sitter solution. %
The solution is 
\begin{equation}
ds^2 = - \left( 1 - \frac{2G M}{r} - \frac{\Lambda r^2}{3}\right)dt^2 + \left( 1 - \frac{2G M}{r} - \frac{\Lambda r^2}{3}\right)^{-1} dr^2 + r^2(d\theta^2 + sin^2 \theta d\phi^2).   
\end{equation}
From this expression one  sees from $g_{00}$ that, since the potential $- \frac{G M}{r}$ gives an attractive force proportional to $- \frac{G M}{r^2}$, the cosmological constant analogously gives an additional potential $ - \frac{\Lambda r^2}{6}$ yielding a 
repulsive force (for $\Lambda>0$)   proportional to $\frac{ \Lambda r}{3}$ and independent of $M$, a force that grows with distance and eventually dominates the attraction from the heavy body of mass $M$.   The comment by Freund, Maheshwari and Schonberg is worth recalling: 
\begin{quote}
 A ``universal harmonic oscillator'' is, so to speak, superposed on the Newton law. The
origin of this extra ``oscillator'' term is, to say the least, very hard to understand. \cite{FMS}
\end{quote}
That description seems to fit $\Lambda<0$ especially, whereas a positive $\Lambda$'s repulsive anti-oscillator potential seems even worse.


\section{Physicists Come to Reject the $\Lambda$-Graviton Mass Conflation}

\subsection{Otto Heckmann (1942) } 

Einstein's  mistake seems to have been noticed first  in 1942 by Otto Heckmann in Germany \cite{Heckmann}.  (For more on Heckmann, see Hubert Goenner's encyclopedia article \cite{HeckmannGoenner}.) 
Previously Heckmann and coauthor Siedentopf had embraced Einstein's  analogy \cite[p. 88]{HeckmannSiedentopf}, claiming that Einstein's equations with a cosmological constant (their equation (5, 17)) has as an approximation $\Delta V + \lambda^2 V = -4\pi \chi \rho,$ their equation (5, 18).     
Heckmann's critique of  Einstein's analogy\footnote{ 
``In diesem Zusammenhang sei folgendes bemerkt: Die von (21) verschiedene
Gleichung $\Delta \Phi + \lambda \Phi  = 4\pi G \rho$ wird von EINSTEIN in der S. 2. Anm. 4 zitierten
Arbeit zur Erl\"{a}uterung der Einf\"{u}hrung des Gliedes $\lambda g_{\mu\nu}$ in seine Feldgleichungen
herangezogen. Dieser bereits von C. NEUMANN gemachte Ab\"{a}nderungsvorschlag
des NEWTONschen Gezetzes (vgl. S. 1 Anm. 1) ergibt sich aber \emph{nicht} als N\"{a}herung
aus den Feldgleichungen der Relativit\"{a}tstheorie. Damit ist die Begr\"{u}ndung,
die HECKMANN und SIEDENTOPF [Z. Astrophys. {\bf 1}, 67 (1930)] f\"{u}r ihre
Gleichung (5, 18) gegeben haben, hinf\"{a}llig.'' \cite[p. 15, emphasis in the original]{Heckmann}  } thus involved a retraction.  This passage 
  is  freely rendered in English by Harvey and Sch{u}cking: 
\begin{quote}  
The equation $\Delta + \lambda \phi = 4 \pi G \rho$ [\emph{sic}:   $\phi$ is missing from the first term] which is different from $\Delta + \Lambda_0(t) = 4 \pi G \rho$ [\emph{sic}:  again $\phi$ is missing from the first term] is used by Einstein in his paper S.-B. Preuss. Acad. Wiss. \emph{\underline{1917}}, 142 to explain the introduction of the term $\lambda g_{\mu\nu}$ into his field equations.  This suggestion for a change of Newton's law
(C. Neumann: ``About the Newtonian Principle of
Action at a Distance,'' p. 1 and 2, Leipzig 1896---see
also Leipziger Ber. Math.-Phys, Kl. 1874, 149) does
\emph{not} result as an approximation of the field equations
of relativity theory. Thus, the argument which Heckmann
and Siedentopf [footnote suppressed] gave for their Eq. (5.18) is
void. \cite[emphasis in the original]{HarveySchucking}.
\end{quote}  
However, given the wartime focus on the practical, the divide between Allied and Axis countries, and the crude Nazi  opposition to Einstein's relativistic physics and so-called `Jewish' physics more generally,  Germany in 1942 was  not an opportune time and place for serious criticisms of some specific aspect of General Relativity to draw worldwide notice. 
Heckmann's book was republished in 1968 but still not widely read \cite{HarveySchucking}.


\subsection{Bryce DeWitt}

In his distinctive  mathematical style, Bryce DeWitt made clear in his 1963 Les Houches lectures (published both in the proceedings \cite{LesHouchesRelativityGroupsTopology} and as a separate book \cite{DeWittDToGaF}) that a cosmological constant is quite distinct from a graviton mass. A graviton mass requires a background geometry, which the cosmological constant does not involve.  He further emphasized the connection between mass terms and smaller symmetry groups.  It was recognized in the late 1930s that whereas massive particles/fields naturally lack gauge freedom, massless particles/fields, at least for spin $1$ and higher, naturally have gauge freedom and correspondingly fewer degrees of freedom \cite{PauliFierz,FierzPauli}.  Thus DeWitt points out that the cosmological constant $\Lambda$  does not shrink the symmetry group, as a real graviton mass term would do, but leaves the general relativistic gauge (coordinate) freedom.  Thus this is  another way to see that $\Lambda$ does not give a graviton mass.  (It is in fact possible to have both a cosmological constant and a graviton mass, but that is another matter, and subtle questions of definition arise.) 

As background for his remarks on gravity (spin $2$), one can consider the simpler and uncontentious case of electromagnetism.  Instead of gauge freedom (as in Maxwell's ``massless photon'' electromagnetism), one has as consequences of the field equations, for the case of massive electromagnetism, in effect the Lorenz-Lorentz condition $ m^2 \partial_{\mu} A^{\mu}$ on the potentials.  Thus the time-like potential $A_0$ is not an independent degree of freedom.  For sufficiently high spins (such as $2$), there is also the possibility of taking the trace and identifying a scalar field within the tensor field content.  If gravity is to be pure spin $2$, then this trace must vanish.  DeWitt contemplates whether these conditions hold on small disturbances, such as waves.

 He   wrote:  
\begin{quote} 
On comparing equation (16.10) [a complicated expression for the second functional derivative of the action for General Relativity] with equation (6.69)  one is at first sight led to  infer that small disturbances in the metric field propagate like those of a tensor
field of rest mass $m = ( -\lambda)^{1/2} $ $(\hbar = c = 16 \pi G = 1)$. This inference is incorrect,
however, for two reasons. In the first place the concept of rest mass requires for
its definition the presence of an asymptotically flat space-time. Indeed space-time
is assumed to be everywhere flat in the linearized theory to which equation (6.69)
refers, whereas, in virtue of (16.9) [Einstein's equations with the cosmological constant], the background field of equation (16.10) cannot be even asymptotically flat. It is true that homogeneous isotropic cosmological
solutions of equations (16.9) exist which can provide a background field with
respect to which a decomposition of small disturbances into positive and negative
frequency components can be effected just as for flat space-time theories. However,
these components necessarily have physical properties which differ to such an
extent from the plane wave components of flat space-time theories that the rest
mass concept is no longer valid. 

In the second place, since the operator (16.10) does not have a form precisely
analogous to that of the tensor field of mass $m$, it does not lead to conditions
$g^{\mu\nu} \delta g_{\mu\nu} = 0,$ $\delta g_{\mu\nu.}^{\; \;\; \;\; \; \;\; \nu}= 0$ on the small disturbances, analogous to the conditions [of vanishing trace and $4$-divergence] of part (c) of Problem 3. In fact, if one attempts to repeat, in generally covariant form,
the arguments which, in the flat space-time theory, lead to such conditions, one
finds in virtue of the lack of commutativity of covariant differentiation, that a
Ricci tensor always appears in such a way as completely to cancel the cosmological
constant via the dynamical equations (16.9). The basic reason for this, of course, is
that the coordinate transformation group is still present as an invariance group for
the gravitational field, and the operator (16.10), despite appearances, is a singular
operator. The time-like components of $\delta g_{\mu\nu}$ are therefore not dynamically suppressed,
and $\delta g_{\mu\nu}$ may, in fact, satisfy four arbitrary supplementary conditions at
each space-time point.
  \cite[pp. 131, 132, footnotes suppressed]{DeWittDToGaF}
\end{quote} 
Whether or not DeWitt had interests in the history of the question, this was a clear, deep, authoritative and twice-published critique of the confusion between a cosmological constant and a graviton mass.  It might have helped that DeWitt straddled the GR-particle physics border.


\subsection{Hans-J\"{u}rgen Treder, Andrzej Trautman, and Wolfgang Rindler}  

Critiques of Einstein's analogy and related ideas from Treder, Trautman, and Rindler are all somewhat similar.  It is noteworthy that in contrast to  DeWitt, these three figures are closer to paradigm general relativists.  

In comparison to DeWitt,  Treder made a more explicitly polemical attack on the idea that the cosmological constant gives gravitons a rest mass \cite{TrederGravitonen,Treder}. The claim that $\Lambda$ implies a graviton mass is a higher-tech version of Einstein's analogy between $\Lambda$ and the modified Poisson equation.    While mentioning ``authors'' he cites the book on unified field theories by Marie-Antoinette Tonnelat \cite{TonnelatUnitaires}. 
Treder introduces Einstein's equations and compares two wave equations, one with the cosmological constant and one with a graviton mass \cite{TrederGravitonen}.  Regarding the equation with the cosmological constant (a zeroth order term in the wave equation), he says:    
\begin{quote}  [t]he constant $\lambda$ cannot be set proportional to  $k^2$, where $ k^{- 1}$ is the Compton wavelength of the graviton. If we form the equivalent of (1.7) [the linearized wave equation without the cosmological constant] to the new equations (1.10) [Einstein's equations with the cosmological constant], we get $$ \frac{1}{2} \Box \gamma_{\mu\nu} + \lambda (\eta_{\mu\nu} + \gamma_{\mu\nu}) = 0. \hspace {1in} (1.11) $$ So that, however, $\lambda$ could in essence be  $k^2$, instead of (1.11) $$\frac{1}{2} \Box \gamma_{\mu\nu} + \lambda \gamma_{\mu\nu} = 0 \hspace {1in} (1.11a) $$ would have to apply'' [references suppressed]     \cite{TrederGravitonen}
\footnote{``Die Konstante $\lambda$ kann nun aber nicht proportional $k^2$ gesetzt werden, wobei $k^{-1}$ die Comptonwellenl\"{a}nge des Gravitons w\"{a}re.  Bilden wir n\"{a}mlich das \"{A}quivalent von (1.7) [the linearized wave equation, without the cosmological constant] zu den neuen Gleichungen (1.10) [Einstein's equations with the cosmological constant], so erhalten wir 
$$\frac{1}{2} \Box \gamma_{\mu\nu} + \lambda(\eta_{\mu\nu} + \gamma_{\mu\nu}) =0.  \hspace{1in} (1.11) $$
Damit aber $\lambda$ im wesentlichen $k^2$ sein k\"{o}nnte, m\"{u}{\ss}te anstelle von (1.11)
$$\frac{1}{2} \Box \gamma_{\mu\nu} + \lambda \gamma_{\mu\nu} =0  \hspace{1in} (1.11a) $$
 gelten'' [references to papers by L. de Broglie and M.-A. Tonnelat]. } \end{quote} 

The contrast between the two wave equations is clear:  the cosmological constant introduces a zeroth order term  $\lambda \eta_{\mu\nu},$ which is more important for weak fields than is the  first order term $\lambda \gamma_{\mu\nu} $ that would signify a graviton mass.  He also mentions how general covariance is retained and hence there are only two wave polarizations even with the cosmological constant.  

In the later paper \cite{Treder}, Treder discusses a tempting mistake with the wave equations that people claiming that a cosmological constant gives a graviton mass sometimes make.  (This is a more sophisticated mistake than Einstein's original one.)
\begin{quote} 
The authors who interpret the cosmological constant $\lambda$ like the
square of the rest-mass of gravitons\ldots, 
put forward as a general argument for their opinion that the variation of the cosmological
equation (1) gives Yukawa-type equations for the perturbations
of the gravitational field. \cite{Treder}
\end{quote}
But this is wrong, as one sees once one keeps track of the covariant derivatives.  
\begin{quote} 
Therefore, the terms of [the weak field wave equation] with the cosmological constant $\lambda$  are
compensated for by the terms with the Ricci tensor.

The result is that the same propagation equations (17) for the
perturbations $\delta g_{kl}$  result from the cosmological equations (1) as from
the equations
$$ R_{kl} = 0.$$
This means that the final form of the propagation equations for the
perturbations of the gravitational field is independent of the existence
of a cosmological term in the Einstein vacuum equations. Therefore,
the gravitons connected with these perturbations have zero rest-mass
for a cosmological constant $\lambda \neq 0$ too. The cosmological constant
does not have any connection with a graviton rest-mass. \cite{Treder} 
\end{quote} 
Thus Treder's and DeWitt's points are basically the same, apart from Treder's explicit polemical aim.

Andrzej Trautman, a very mathematical general relativist, accurately critiqued the $\Lambda$-graviton mass confusion in some book-length lectures at Brandeis in  1964, along with a brief passable reference to the history \cite[pp. 228-231]{Trautman}.  After discussing Olbers' paradox and the problem of the diverging Newtonian potential for an infinite homogeneous matter distribution, Trautman  says: 
\begin{quote} 
Neumann and Seeliger (in 1895) proposed the idea of  replacing Poisson's equation by
$$ \nabla^2 \phi - \lambda \phi = 4 \pi k \rho. $$
This corresponds to assuming that the gravitational forces
have a finite range with $1/\sqrt{\lambda}$ being the characteristic length for the gravitational interactions.\ldots

\ldots Einstein modified the field equations by adding a cosmological term 
$$ \lambda g_{ab} + R_{ab} - \frac{1}{2} g_{ab}R = -8 \pi k T_{ab} \hspace{1in} (9.2) $$
However, these equations are not the analog of the Neumann-Seeliger equation in the Newtonian limit but go over into 
$$ \nabla^2 \phi + \lambda c^2 = 4 \pi k \rho. $$ 
\cite[pp. 229, 230]{Trautman} 
\end{quote}  
While not linking the Neumann-Seeliger 19th century idea to the 1920s-30s physical meaning of a graviton mass or being explicitly interested in Einstein's own making of the analogy, Trautman nonetheless makes clear the distinction between the two things that Einstein had presented as analogous.  




Wolfgang Rindler also gently  criticized the analogy between $\Lambda$ and the exponential decay of Neumann and Seeliger \cite{RindlerFirst}. The problem of the infinite potential in a homogeneous matter distribution in absolute space
\begin{quote}
\ldots  led Neumann and Seeliger in 1896 to suggest that the
Newtonian potential of a point mass be replaced by
$$ \phi = - \frac{mG}{r} e^{-r\sqrt{\lambda}}, \hspace{.3in}  (\lambda = constant \approx 0), \hspace{1in} (82.1)$$
whose integral would remain finite. (Note that this is
identical in form with Yukawa's mesonic potential put forward
in 1935.) Poisson's equation [footnote suppressed] $ \nabla^2 \phi = 4 \pi G \rho$ then becomes
$$ \nabla^2 \phi - \lambda \phi = 4 \pi G \rho, \hspace{1in} (82.2) $$
which possesses the obvious solution
$$ \phi = - \frac{4 \pi G \rho}{\lambda} \hspace{1in} (82.3)$$
in a homogeneous universe. (This results also on integrating
(82.1) throughout space for a continuous distribution
of matter.)

It is interesting to observe the striking formal analogy
between Einstein's modification (79.10) of his original field
equations (79.7) and Neumann and Seeliger's modification
(82.2) of Poisson's equation. However, in first approximation
(79.10) does not reduce to (82.2) but rather to
another modification of Poisson's equation, namely
$$ \nabla^2 \phi + c^2 \Lambda = 4\pi G \rho, \hspace{1in} (82.4) $$
as can be shown by methods similar to those of Section 79.
This \emph{also} admits a constant solution in the presence of
homogeneous matter, namely $\phi = 0$, \emph{provided} $c^2 \Lambda = 4\pi G \rho$---a situation which obtains in Einstein's static universe, for which, indeed, Einstein originally introduced his $\Lambda$ term. 
\cite[pp. 222, 223]{RindlerFirst} \end{quote}
Though the contributions of Neumann and Seeliger are run together, this description in an accessible textbook is both historically and technically serviceable.


\subsection{Peter G. O. Freund, Amar Maheshwari, and Edmond Schonberg}

One of the clearest distinctions between a cosmological constant and a graviton mass comes from a significant particle physics-flavored paper (in the \emph{Astrophysical Journal}!) putting forward a theory of massive gravitons \cite{FMS}.
Einstein's theory  with the cosmological constant is faulted on multiple grounds, of which here is the second. 
\begin{quote} 
B.  In the ``Newtonian'' limit it leads to the potential equation,
$$ \Delta V + \Lambda = \kappa \rho . \hspace{1in} (1) $$
Correspondingly, the gravitational potential of a material point of mass $M$ will be given
by 
$$ V = -\frac{1}{2} \Lambda r^2 - \frac{\kappa M}{r} . \hspace{1in} (2) $$
A ``universal harmonic oscillator'' is, so to speak, superposed on the Newton law. The
origin of this extra ``oscillator'' term is, to say the least, very hard to understand. \cite{FMS} \end{quote} 
By contrast, their proposed massive graviton theory is far more reasonable by the standards of particle physics, and because it violates general covariance, it is easier to quantize as well.\footnote{They were presumably not reckoning sufficiently with the ghost problem, though they did discuss it.  But their appendix is evidently the first public appearance of a \emph{nonlinearly} ghost-free, that is pure spin $2$, massive gravity, singled out from among the OP $2$-parameter family of theories.  The nonlinear \emph{argument} was published later \cite{MaheshwariIdentity} (submitted no later than early April 1971).  This the (Tyutin-Fradkin-)Boulware-Deser nonlinear ghost problem was pre-solved before it was proposed. But no one noticed and the problem had to be solved again in 2010.  Thus the decades of darkness for massive gravity were quite contingent. Maheshwari was unaware of the van Dam-Veltman-Zakharov discontinuity, however.}
\begin{quote} 
In the Newtonian limit, equation (1) is now replaced by the Neumann-Yukawa
equation,
$$ (\Delta -m^2) V  = \kappa \rho , \hspace{1in} (3) $$
which leads to the quantum-mechanically reasonable Yukawa potential,
$$ V(r) = - \frac{\kappa M e^{-mr} }{r} , \hspace{1in} (4)$$
rather than the peculiar oscillator of equation (2). Difficulty B is thus removed. 
\end{quote} 
Thus during the 1960s quite a few leading physicists took aim at the confusion between the cosmological constant and the graviton mass, confusion that apparently had gone  unchallenged apart from Heckmann's little-noticed  work. %


\section{Engelbert Schucking's Decisive Influence}

While some authors since the 1960s have criticized Einstein's conflation of his cosmological constant with a graviton mass, 
Engelbert Schucking (also spelled ``Sch\"{u}cking'') later mounted a sustained assault on that error  \cite{Schucking,HarveySchucking}.  
Sch{u}cking's first work appeared in a \emph{Festschrift} for Peter Bergmann, a work likely to be read only by physicists, though the content is substantially historical.  Sch{u}cking's work on the Einstein Papers project also implies that one could as plausibly list his contribution among the historians and philosophers to be treated later as among the physicists treated above,  though his work somehow  had no influence on the treatment of the  analogy in the Einstein Papers.   Given the difficulty of classifying him and the transformative nature of his  interventions, it seems fitting to devote a separate section to his influence.  

His first writing  is so accurate, brief, and vigorous that it is tempting to quote the whole thing, though one might wish that Seeliger got mentioned along with Neumann.  Here is a substantial portion. 
\begin{quote}
To motivate the introduction of this new constant of nature without a wisp of empirical evidence
he wrote that his new extension was ``completely analogous to the extension of the Poisson equation
to 
$$ \hspace{1.5in} \Delta \phi - \lambda \phi = 4\pi K \rho \hspace{.5in}   \mbox{''}    \hspace{1.6in}  (3)$$

This remark was the opening line in a bizarre comedy of errors.

Einstein's modified Poisson equation is now familiar to all physicists through the static version
of Yukawa's meson theory which has the spherically symmetric vacuum solution
$$ \hspace{.75in} \phi = \frac{const}{r}  e^{-r \sqrt{\lambda} }    ,  \hspace{.3in}  \lambda = (\frac{mc}{\hbar})^2  , \hspace{.3in}  r = (x^2 + y^2 + z^2)^{\frac{1}{2}} .  \hspace{.4in}  (4)$$

But this equation had a deeper root. The K\"{o}nigsberg theoretician Carl Neumann (Neumann
1896) had proposed the modified Poisson equation to introduce an exponential cut-off for the
gravitational potential. He thus anticipated Einstein's worry about the disastrous influence of distant
stars on the potential. Einstein, apparently, was not aware of Neumann's work in 1917.

It is true that the Poisson equation modified by a term $-\lambda \phi$ (with a positive $\lambda$ ) on its left
hand side leads to an exponential cut-off for the gravitational potential. But Einstein's flat assertion
that the $\lambda$ -term in his field equations had a completely analogous effect was wrong. However
generations of physicists have parroted this nonsense. Even Abraham Pais (Pais 1982) writes in
his magisterial Einstein biography about the analogy between the $\lambda$-terms in Poisson's and
Einstein's equations ``he (Einstein) performs the very same transition in general relativity''.

It seemed so deceptively obvious: the potential corresponds in the Newtonian approximation
to ($c = 1$)
$$ \hspace{2in} g_{00} =-(1+2\phi).    \hspace{1.7in}   (5) $$

Thus adding a term $-\lambda \phi$ to $\Delta \phi$ might correspond to inserting a term $-\lambda g_{\mu\nu}$  in addition to the
Ricci tensor whose $00$-component gives essentially the Laplacean in Newtonian approximation.


I still remember when Otto Heckmann told me 35 years ago: ``Einstein's Argument ist
naturlich Quatsch (baloney)''.  And the late Hamburg cosmologist was right.  For $\phi$ is $\phi / c^2$  and
can be neglected compared to one in first approximation. Thus the Newtonian analog of
Einstein's equations with $\lambda$-term is not the modified Poisson equation (3) but
$$ \hspace{2in} \Delta \phi + \lambda c^2 =4\pi K \rho.   \hspace{1.6in}    (6) $$
With equation (6) Einstein had not introduced an exponential cut-off for the range of gravitation
but a new repulsive force ( $\lambda > 0$), proportional to mass, that pushed away every particle of
mass $m$ with a force

$$ \hspace{2in} \vec{F}  = m c^2 \frac{\lambda}{3} \vec{x},  \hspace{2in}     (7) $$  

a force derivable from the repulsive oscillator potential $-\lambda c^2 r^2 /6 $.  This was clearly stated by Arthur Eddington (Eddington 1923).\ldots

Instead of getting a shielded gravitation one had now at large distances almost naked repulsion.
This was quite different from the expected bargain.
\cite[pp. 185, 186]{Schucking}
\end{quote}
This Italian \emph{Festschrift} for Peter Bergmann was probably a bit too obscure and technical to reach the widest relevant audience, however.  
As perhaps a foretaste of how old errors die hard, Joe Weber in the same volume \cite{Weber} made the same mistake that Sch{u}cking corrected!  At least Weber was  thinking about experimental tests of a graviton mass,  calling attention to Zwicky  \cite{ZwickyGravitonMass}, who was influenced by a Caltech colleague  Feynman. 
One might think that Sch{u}cking's work has already said everything that needs saying.  But the persistence of unclear and even erroneous ideas in the newer historical literature, after both the 1960s physics corrections and in  some cases after Sch{u}cking's blasts, shows that the topic requires continuing discussion.  
Doubtless his own personal connection to Heckmann played a role in his work on this topic. 
A more recent article by Alex Harvey and Engelbert Schucking \cite{HarveySchucking}, published in a more visible and pedagogical place (\emph{American Journal of Physics}), takes much the same message (with a fair amount of reused text) to a larger audience.

%


\section{History and Philosophy of Science  and Einstein's  Analogy }  

Unfortunately, historical treatments of Einstein's cosmological constant $\Lambda$ and its relevance to the Seeliger-Neumann modification of Newtonian gravity  have not always been  reliable. That is despite the fact that some of those here classed as historians were trained as and long operated as physicists; their classification as historians here reflects more their high achievements in history than any dearth of participation in physics.  
 This section might be an interesting case study on the need for an at least partly internalist history of science, in  that a purely externalist view would not be motivated to inquire further to trace the origins of erroneous claims.  It also helps to illustrate how systematic neglect of particle physics by historians and philosophers of General Relativity leads to errors and oversights.  
  

\subsection{John D. North}

John D. North addressed the issues in question on more than one occasion in his historically rich work.  His classic 1965 work starts well in its discussion of Seeliger and Neumann, but later tends to confuse their idea with that of Laplace and with Einstein's cosmological constant.  

\begin{quote}
In 1895 Seeliger began by protesting that as the volume $(V)$ of a
Newtonian distribution of matter of finite density tends to infinity, the
gravitational potential at any point 
 can be assigned no definite value; added to which the expression for the gravitational force also becomes
indefinite. Carl Neumann, faced by the same difficulties, proposed that
Poisson's equation should be adjusted so as to permit a uniform and
static distribution of matter throughout space. For the gravitational
potential they took expressions of the usual Newtonian form, multiplied
by an additional factor $e^{-\alpha r}$, where $\alpha$ is a quantity sufficiently small to
make the modification insignificant, except for large distances.\ldots 

\ldots    On the other hand, neither were Seeliger and Neumann first with the exponential law: Laplace had
taken this very law fifty years before. 
  In all the earlier cases, however,
the concern was in only a narrow sense cosmological.

 The effect of the exponential modifying factor is to introduce a
cosmical repulsion capable, at large distances, of exceeding the usual gravitational forces. As will be seen in due course, the introduction of  what was to be known as the `cosmological term' into the later gravitational field equations of Einstein is reminiscent of Neumann's modification
of Poisson's equation. \cite[pp. 17, 18, footnotes suppressed]{North}
\end{quote}
Here Seeliger's diversity of mathematical expressions has been pared down to match the more unique and optimal expression of Neumann.  
North's effort  to find an antecedent in Laplace \cite[book 16, chapter 4, p. 481]{Laplace}, unfortunately, confused Laplace's multiplication of a $\frac{1}{r^2}$ \emph{force} by an exponentially decaying factor with Neumann's multiplication of the $\frac{1}{r}$ \emph{potential} by such an exponential factor.  (The same mistake was made by Erich Robert Paul  \cite[p. 69]{SeeligerStatistical}.)  
 Laplace's modified force law is not obviously the solution to any relevant linear differential equation or connected with any clear physical meaning of current interest, so Neumann's exponential decay modifying the potential is more plausible.  
And North is in mathematical error in holding that the  exponential modifying factor introduces ``a
cosmical repulsion capable, at large distances, of exceeding the usual  gravitational forces.'' The exponential decay merely causes the gravitational attraction to weaken faster than it otherwise would.  He  seems to  be warming up for confusing Neumann's exponentially decaying factor in the potential (or its ancestor in the relevant differential equation, the modified Poisson equation) with Einstein's cosmological constant $\Lambda.$

North unfortunately makes a version of Einstein's erroneous conflation in discussing the Milne-McCrea modified Newtonian cosmology that sought to encompass much of relativistic cosmology within a simpler framework.  This work permitted a $\lambda$ term, an introduction by hand of a term like Einstein's cosmological constant $\Lambda$ into Newtonian equations \cite{McCreaMilneNewtonian}. According to North, 
\begin{quote}
(The $\lambda$-term in the equation of motion is precisely that which Neumann and Seeliger had introduced, nearly forty years before, in the hope of explaining how an infinite and static universe was possible.)  \cite[p. 179]{North}  
\end{quote}

North's more recent work seemed not to profit fully from the somewhat greater visibility of critiques of Einstein's $\Lambda$ error that had become available.
\begin{quote}
When applied to cosmological problems assuming an infinite Universe, ordinary Newtonian theory, based on the familiar (Euclidean) geometry, seemed to lead to inconsistencies.  In fact Carl Neumann and Hugo von Seeliger --- to name only two --- tried to modify the Newtonian law of gravity to remove these difficulties.  In doing so, strangely enough, they introduced what was effectively a
cosmical repulsion (one that they supposed worked against the much more powerful gravitational attraction), which has its counterpart in later relativistic cosmology.  \cite[pp. 515, 516]{NorthNorton}  \end{quote}
%


\subsection{Max Jammer and Abraham Pais} 

Max Jammer's generally impressive and erudite work has, unfortunately, also tended to perpetuate Einstein's confusion about $\Lambda$  \cite{JammerSpace,JammerMass}. 
 Jammer wrote in \emph{Concepts of Space}:
\begin{quote}
Einstein's introduction of the cosmological constant $\lambda$, by which he hoped to remove the inconsistency with Mach's Principle, stands in striking similarity to H. Seeliger's modification of the classical Laplace-Poisson equation $\Delta \varphi = 4 \pi \rho$ into $\Delta \varphi - \mu^2 \rho = 4 \pi \rho$, [\emph{sic}---the left side should contain $\varphi$ rather than $\rho$] whereby Seeliger attempted to relieve Newtonian cosmology from certain inconsistencies. 
 The positive constant $\mu$ should be chosen so small that within the dimensions of the solar system the solution of the original equation (i.e., $\varphi = -m/r$) and that of the supplemented equation ($\varphi = - \frac{m}{r} e^{-\mu r}$) should coincide within the margin of observational error.  


 \cite[pp. 194, 195, footnotes suppressed]{JammerSpace}.  \end{quote}
(There is no change from  the 1969 edition.) 
While criticisms of Einstein are admitted, there seems to be no expectation that they would come from much later decades (1940s-60s) and be facilitated by particle physics.  
Notwithstanding the citation of Seeliger's 1895 and 1896 papers, Jammer's Seeliger looks more like Neumann (who does not appear here in Jammer's account) or a composite figure propounding a fusion of Seeliger's physical arguments and Neumann's mathematics in presenting a unique plausible differential equation, rather than a variety of more or less arbitrary modified force laws as Seeliger in fact gave.  



More recent  historical works show no monotonic  improvement.  Pais's work is similar to Jammer's on this point; thus Einstein's standard scientific biography  did no better in the 1980s, well after the  corrections were  available in the physics literature, than Jammer had in the 1960s.  Pais discusses Einstein's proposed modification of Newtonian gravity well enough, but then endorses Einstein's analogy. If 
\begin{quote} the Newton-Poisson equation $$ \Delta \phi = 4 \pi G \rho \hspace{1.5in} (15.17)$$\ldots
 is replaced by $$  \Delta \phi -  \lambda \phi = 4 \pi G \rho \hspace{1.1in} (15.18)$$ (a proposal which again has nineteenth century origins), then the solution [with uniform matter density $\rho$ and gravitational potential $\phi$] is dynamically acceptable.

\ldots 
Let us return to the transition from Eq. 15.17 to Eq. 15.18.  There are three main points in Einstein's paper.  First, he performs the very same transition in general relativity\ldots 
\cite[p. 286]{PaisEinstein}. \end{quote}


\subsection{Pierre Kerszberg} 

In a generally fascinating and illuminating book written in the late 1980s,  Kerszberg seems to have lacked  the concept of a graviton mass and consequently had an obsolete view of the plausibility of a Neumann-Seeliger modification of the Newtonian potential, while also falling into Einstein's false analogy. (Schucking commented on Kerszberg's book \cite{Schucking}).  Kerszberg writes, 
\begin{quote}
By the end of the nineteenth century, Seeliger and Neumann proposed a modification of the Newtonian law of gravitation that would make no perceptible difference within the solar system but would dispense with the disturbing increase of attraction over larger distances.\ldots 
 Suffice to say here that this modification was designed to restore homogeneity at large, and that it was just as \emph{ad hoc} a solution as the absorbing matter had been in the case of the optical paradox. \cite[p. 50]{KerszbergInvented}.
\end{quote}
The promised further discussion in the next chapter largely embraces Einstein's analogy. 
\begin{quote}
	In fact Einstein introduces the cosmological constant by again appealing to a parallel with the Newtonian theory.  As he reminds us at the end of the analysis of Newtonian cosmology with which he opened his memoir, the strategy for overcoming the paradoxes of the island universe involves what looks like a piece of similar contrivance.  It was in the third edition of his popular exposition of relativity that Einstein gave Seeliger credit for the modification of Newton's law, according to which ``the force of attraction between two masses diminishes more rapidly than would result from the inverse square law'' (1918\ldots).  In fact, C. Neumann had reached similar conclusions (see Seeliger 1895 and Neumann 1896).  Einstein went on to emphasize that neither a theoretical principle nor an observation would ever justify the proposed modification; any other convenient law would do the same job.\ldots Seeliger's modification of the inverse square law was $F=Gmm^{\prime} e^{-\Lambda r}/r^2.$ \cite[pp. 161, 162]{KerszbergInvented} \end{quote}
Unfortunately this expression of Seeliger's (one of a number that he used) was not the most plausible option even in the 1890s.  Neumann's potential had the virtue of solving a known linear differential equation, unlike most of Seeliger's expressions. 
The enhanced plausibility of Neumann's mathematics with the rise of the concept of graviton mass overcomes the objections to Seeliger.  
Kerszberg commendably finds $\Lambda$ difficult to interpret. 

He comments more explicitly on Einstein's analogy between his modified Poisson equation and his cosmological constant $\Lambda$:
\begin{quote} 
Beyond any surface similarity, there is indeed a fundamental contrast, because the $\Lambda$-term now fixes the periphery
and removes all reference to the centre. Thus, the new constant is parallel to
Newton insofar as the \emph{form} of the laws is concerned, but the \emph{interpretation} of
it diverges sharply from formal analogy. \cite[p. 163]{KerszbergInvented} \end{quote} 
But in fact scalar (spin $0$) and symmetric tensor (spin $2$) tensor theories of gravity are too similar to permit this disconnect between formal analogy and interpretation \cite{FMS,DeserMass}.  In fact there is no formal analogy, either; Kerszberg's sense of an important difference is correct.  
 Once one has the proper alignment of scalar/spin $0$ and tensor/spin $2$ analogs, the behavior of the corresponding cases (scalar with graviton mass, tensor with graviton mass; scalar with cosmological constant, tensor with cosmological constant) is quite similar, both empirically and conceptually.  In both the scalar and tensor cases, the massive graviton case involves a background geometry and a smaller symmetry group (typically the Poincar\'{e} group of special relativity), whereas the cosmological constant involves no background geometry and does not shrink the symmetry group.

%
%


\subsection{John Norton and John Earman:  Clarity But No Graviton Mass}

Some philosophically flavored history out of Pittsburgh has fared noticeably better.  At the end of the 1990s, John Norton very helpfully discussed ``The Cosmological Woes of Newtonian Gravitational Theory,'' providing a  useful discussion of Seeliger, Neumann, and Neumann's elusive priority claim.  Norton, citing Trautman, is not taken in by Einstein's claimed analogy between the modification of Newtonian gravity and the cosmological constant:
\begin{quote} 
(The analogy is less than perfect---something Einstein may have
found it expedient to overlook in order not to compromise his introduction of the
cosmological term. As Trautman (1965: 230) pointed out, Einstein's augmented
gravitational field equation reduces in Newtonian limit to a field equation other
than [the modified Newtonian equation $\nabla^2 \phi - \lambda \phi = 4 \pi G \rho]$:  $ \nabla^2 \phi + \lambda  c^2 = 4 \pi G \rho.$)
\cite[p. 299]{NortonWoes} \end{quote} 
  Norton does not, however, consider how developments in the 1920s-30s in particle physics provided a physical meaning for Neumann's mathematics and thus made a specific and plausible form of Seeliger's ideas more available.  Admittedly, changing gravitation on long ranges becomes far less urgent once one finds that the universe is expanding. 

John Earman's paper on $\Lambda$ \cite{EarmanLambda}, which draws upon Norton's, is also clear and accurate technically---indeed clearer than others that say little about the perturbative business involved---as well as historically aware.  He does not, however,  connect the Neumann-Seeliger work with the later physical meaning of a graviton mass.  


\subsection{Helge Kragh: Rewriting History} %

More recently, in an otherwise admirable  book \emph{Matter and Spirit in the Universe:  Scientific and Religious Preludes to Modern Cosmology}, Helge Kragh took Einstein's false analogy between $\Lambda$ and the Seeliger-Neumann sort of modification of gravity so seriously as to rewrite history in light of this analogy, at variance with what Seeliger actually proposed \cite[p. 28]{Kragh}.  Kragh seems to have used both Einstein's false analogy and the 1918-19 Kottler-Weyl (a.k.a. ``Schwarzschild-de Sitter'') \emph{solution} of Einstein's equations with $\Lambda$ discussed above to infer the mathematical potential comparable to $\frac{1}{r}$ that Seeliger supposedly posited in the 1890s.  This error seems not to appear in Kragh's sources \cite{NortonWoes,HarrisonNewtonInfinite,JakiSeeliger}. 
 Kragh writes:
\begin{quote}
Seeliger's suggestion was to replace the [Newtonian gravitational potential $\varphi(r)=-M/r$]\ldots with 
\begin{eqnarray*}
\varphi(r) = - \frac{M}{r} - \frac{\Lambda r^2}{6} 
\end{eqnarray*}
where $\Lambda$ is a cosmological constant so small that its effects will be unnoticeable except for exceedingly large distances.  The body not only experiences an attractive inverse-square force towards the central body, but also a repulsive force given by $\Lambda r/3$.  In a somewhat different way the slight but significant adjustment of the inverse square law was suggested also by Carl Neumann in 1896, and it reappeared in a very different context in 1917, now as Einstein's famous cosmological constant. \cite[p. 28]{Kragh}.
\end{quote}
Seeliger did experiment with various ways to modify the Newtonian potential \cite{Seeliger1895a,Seeliger1896,Seeliger1898ab} 
and did not seem captivated by Neumann's specific proposal to include just an exponentially decaying factor in the potential (a modification later interpretable as a graviton mass and hence the `right' way to do it). But adding a divergent quadratic potential was not among the things that Seeliger considered. 

%


\subsection{Marco Mamone Capria: Clarity But No Graviton Mass}

The faulty analogy has been treated better  in the work of Marco Mamone Capria \cite{MamoneCapriaLambda}, partly with influence from Sch{u}cking.  Regarding Einstein's proposed modification of Newtonian gravity, it 
\begin{quote}
 was probably inspired by the modified gravitational potential $\phi_N$ built out of the mass-point potential $A e^{-r \sqrt{\lambda}}/r$ ($A$ is a constant), as proposed by the German theoretical physicist Carl Neumann in 1896. Neumann introduced this form of
the potential in order to solve the gravitational paradox in the form of the impossibility
to assign, at any point, a finite value to the gravitational potential corresponding to a
uniform infinite mass distribution. In fact $\phi_N$ goes to zero at infinity even with such a
mass distribution, and (2) [the modified Poisson equation] is precisely the equation which it satisfies. \cite[p. 131]{MamoneCapriaLambda}
\end{quote} 
Seeliger's motivation seems to be attributed to Neumann, however.  
Mamone Capria explicitly rejects Einstein's analogy \cite[p. 135]{MamoneCapriaLambda}, following Harvey and Schucking \cite{HarveySchucking}.
\begin{quote} 
Even more remarkable [than Einstein's false analogy,  the quarter of a century before it was criticized by Heckmann, and the ease with which one refutes the analogy], perhaps, is that a long sequence of eminent authors missed this basic point and blindly endorsed Einstein's stated analogy between (2) and (7) (``generations
of physicists have parroted this nonsense'' as is said in [\cite[p. 723]{HarveySchucking}]). Clearly
most working scientists are just too anxious to publish some `new' piece of research of
theirs to spend a sufficient amount of time reviewing the foundations of their disciplines;
so they frequently end up by relying on authority much more than on rational belief, in
contrast with the scientific ethos as ordinarily proclaimed.
\cite[p. 135]{MamoneCapriaLambda}  \end{quote}
Mamone Capria does not go on to draw the further conclusion that the actual analog of the modified Poisson equation is thus a potentially interesting and unexplored possibility for relativistic cosmology or  recognize that particle physics is the missing ingredient, but still this is progress.


\subsection{O'Raifeartaigh, O'Keeffe, Nahm and Mitton: Clarity  But No Graviton Mass}

A recent centennial review of Einstein's 1917 cosmological paper also makes a clear distinction between the Neumann-Seeliger enhanced long-range decay and Einstein's cosmological constant.  
\begin{quote} 
 It is an intriguing but
little-known fact that, despite his claim to the contrary, Einstein's modification of the
field equations in Section §4 of his memoir was not in fact "perfectly analogous" to
his modification of Newtonian gravity in Section \S1. As later pointed out by several
analysts [footnote suppressed], the modified field equations (E13a) do not reduce in the Newtonian limit
to the modified Poisson's equation (E2), but to a different relation given by
$$ \nabla^2 \phi  + c^2 \lambda = 4\pi G \rho. \hspace{1in} (10)$$
This might seem a rather pedantic point, given that the general theory allowed
the introduction of the cosmological constant term, irrespective of comparisons with
Newtonian cosmology. Indeed, as noted in Section 4.1, Einstein described his modification
of Newtonian cosmology merely as a ``\emph{foil for what is to follow}''. However, the
error may be significant with regard to Einstein's interpretation of the term. Where
he intended to introduce a term to the field equations representing an attenuation
of the gravitational interaction at large distances, he in fact introduced a term representing
a very different effect. Indeed, the later interpretation of the cosmological
term as representing a tendency for empty space to expand would have been deeply
problematic for Einstein in 1917, given his understanding of Mach's Principle at the
time. Thus, while there is no question that relativity allowed the introduction of the
cosmic constant term, it appears that Einstein's interpretation of the term may have
been to some extent founded on a misconception (Harvey and Schucking 2000, p. 223).
\cite[p. 28]{ORaifeartaighLambda}
\end{quote} 
This is a very appropriate assessment, albeit without  seeing that the Neumann-Seeliger idea provided the mathematical nucleus of a potentially interesting case of unconceived alternatives in the form of a graviton mass.


\section{Conclusion}

It seems that there is an irregular but gradual upward trend in recognition by historians and philosophers of physics of the distinction between faster long-range decay  and a cosmological constant, notwithstanding Einstein's claimed analogy.  Thus it is certainly possible to get the correct answer without much personal awareness of particle physics. However, it is also quite easy to fall for the false analogy.  Much depends on what one has read.  Works that clarify the issue are not standard reading material for historians or philosophers of physics.

But there are a number of imperfections even in a situation in which historians and philosophers recognize the erroneous quality of Einstein's analogy.  First, it is unclear that one can  make sense of the difference between a cosmological constant and a graviton mass if one is averse to particle physics; one cannot  make sense of a graviton mass without a background geometry and perturbative-looking comparison of two geometries.  Second, it appears that the increased recognition of  the distinction  is based largely on authority, oftentimes that of Sch\"{u}cking, making the recognition highly contingent on reading the right materials.  Third,  if one does manage to avoid the false analogy, one can easily still fail to recognize that one has encountered a potentially interesting example of the problem of unconceived alternatives.  A dose of particle physics egalitarianism and corresponding de-privileging of General Relativity exceptionalism would help on all counts.

This discussion is clearly not intended to imply that anyone should embrace a graviton mass or should not embrace a cosmological constant.  Ultimately the world should settle that issue through observational data.  At present at least an effective cosmological constant has empirical support, whereas the graviton mass does not clearly have any empirical support (although it is difficult to say what to make of the dark matter problem, which might be soluble in terms of modified gravity in some form).  But failing even to conceive of a graviton mass, which is a plausible option from the standpoint of particle physics, while taking with great seriousness a cosmological constant even prior to any data, when it  is at least \emph{prima facie} implausible from the standpoint of particle physics (prior to \emph{quantum} field calculations), is hardly a balanced view.  One needs to adjust one's prior plausibilities in order to profit more from the data.

The briefly mentioned  examples of gravitational radiation and particle physicists' spin $2$ derivations of Einstein's equations also illustrate how influence from particle physics can be salutary for understanding and motivating Einstein's equations.  All three examples provide support for the claim that systematic integration of general relativist and particle physics ideas holds considerable promise for historical and philosophical reflection on gravitation and space-time.


\section{Acknowledgments}   

Many thanks are due to Alexander Blum  for assistance with translations, editing, points of emphasis, and acquainting me with  Bryce (Seligman) DeWitt's dissertation; to J\"{u}rgen Renn, Roberto Lalli and the rest of the  Discussion Group on the Recent History of General Relativity at the Max-Planck-Institut f\"{u}r Wissenschaftsgeschichte for the opportunity to participate; to Michel Janssen and Dennis Lehmkuhl for discussion; to Karl-Heinz Schlote for help in finding Neumann's works; and to Cormac O'Raifeartaigh for discussion and mentioning the Mamone Capria paper.  This work was supported by the John Templeton Foundation grants \#38761 and \#60745 and the  National Science Foundation (USA) STS grant \#1734402; all views are my own.

%
\end{document}